# THE MANNHEIM-KAZANAS SOLUTION, THE CONFORMAL GEOMETRODYNAMICS AND THE DARK MATTER

M.V. Gorbatenko, S.Yu. Sedov

Russian Federal Nuclear Center - All-Russia Research Institute of Experimental Physics, Sarov, Nizhny Novgorod Region

**Abstract**

Within the framework of the Einstein's standard equations of the general theory of relativity, flat galactic rotational curves of galaxies cannot be explained without hypothesis attracting the dark matter, the particles of which had not yet been identified. The vacuum central-symmetric solution of the equations of conformal gravitation is well known as metrics of Mannheim-Kazanas, on the basis of which these curves receive purely geometrical explanation. We show in our work that the metrics of Mannheim-Kazanas is the solution of not only Bach equations received from conformal-invariant Weyl Lagrangian, but also the solution of equations of the conformal geometrodynamics at a nonzero vector of Weyl. In this connection the hypothesis is formulated that the space on galactic scales can be described not only by Riemannian geometry, but with geometry of Weyl.

## 1 INTRODUCTION

It is well known that the mystery of the dark matter (DM) is still not solved. Its existence is confirmed by numerous observation data that in particular refer to the flat galactic rotational curves. The situation seems as if inside visible (light) matter dark matter is spread in the volume of galaxies, the presence of which reveals itself in not only additional gravity force that affect the visible matter.

Rotational curves of the rotation of stars, hydrogen and other types of light matter around the center of the spiral galaxy starting with the radius of about $1.2R_0$ has a flat view, as they say, i.e. the orbiting velocities in a wide range of radii do not decrease with the distance. Here, $R_0$ is the thickness of the optical disc of the galaxy. The rotational curve for galaxy NGC7331 is shown in Figure 1 adopted from [1]. Other galaxies have similar rotational curves (see [1]-[5], for example).

It is well-known [1]-[6] that the rotational curves are described well by the empirical dependency of the rotational velocity of stars $v$ on the radius that is written as follows

$$v^2 = \frac{aN}{R} + bNR + cR, \qquad (1)$$

where $R$ is the distance to the center of the galaxy, $N$ is the number of stars in the galaxy, and $a$, $b$, $c$ are phenomenological parameters. This formula besides the first Newtonian term has other terms that are proportional to the radius. These terms do not come from the general theory of relativity (GTR) by Einstein, but they give the flat shape of rotational curves.

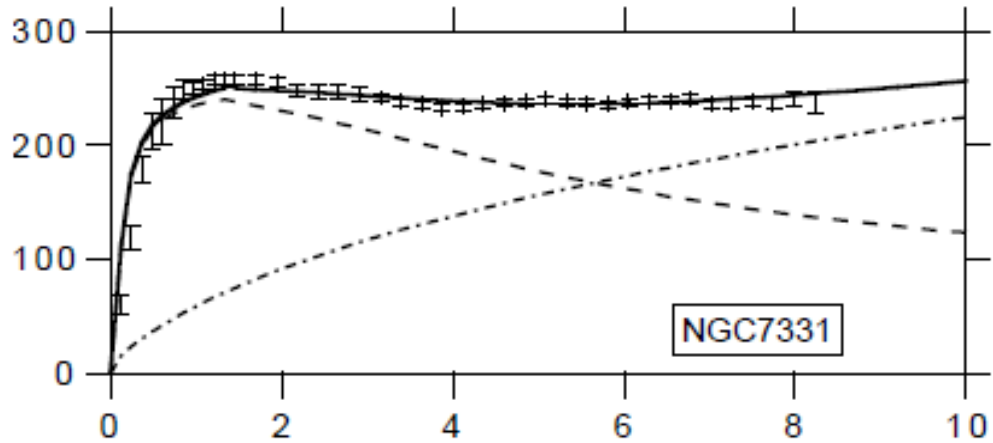

Figure 1 - The orbiting velocities of the visible matter (in km/s) around the center of galaxy NGC7331 as a function of ratio $R/R_0$, where $R_0$ is the thickness of the optical disk, $R$ is the distance from the center of the galaxy. Dashed line shows the contribution of the Newtonian component. Dot-dash line is the contribution of the linear potential. Solid line is the accumulation curve. The data are from [1].

To explain the presence of flat sections, there were used two kinds of hypothesis. The most popular is the hypothesis based on the presence of dark matter (DM) assumption. Within this theory they are looking for the DM carriers. There are many theories considering the nature of the DM carriers [6]; however, none of them has not reached the experimental confirmation yet. This fact makes us look for some alternative ways to explain flat rotational curves. The other alternative search area is in fact the geometric one, i.e. the one when the observed effects (flat rotational curves in this case) are the result of the solution of the equations of some sort modified gravitation theory. The progress in the direction of the geometrical explanation of the rotational curves is done in the work by Mannheim-Kazanas [7], where an exact solution of the conformal gravitation equation is obtained using Weyl Lagrangian. It appeared that space metric in this solution has a term that is proportional to the radius, which is capable to explain flat galactic rotational curves in purely geometric way, without the hypothesis of the dark matter existence. The dynamic equations of space here should be written not in the form of standard GTR equations, but in a more general form – as gravitation equations generalized with the requirements of the conformal invariance. One of these ways of such generalization is to use of the Ladrangian quadratic by Weyl tensor in Riemann space. We call the resulted equations Bach equations in compliance with [8]. That was the exact solution of Bach equations in vacuum that was obtained in [7].

Quadratic variant of conformal generalization of GTR equations is not the only possible one. In [9], [10] a more simple variant of such generalization was offered in the Weyl space. The governing equations in [10] were called the equations of conformal geometrodynamics (CG). They had the view of standard GTR equations, but with a specific tensor of energy-impulse constructed from a new geometric object $A_\alpha$, which we call a Weyl vector. In case when Weyl vector is brought to the gradient from some scalar field, CG equations coincide with Brans-Dicke equations [11] when parameter $\omega = -3/2$. If we add the energy-impulse tensor of the matter to these equations, these equations will comply with the equations proposed in [12] as a scale-invariant gravitation theory.

It is interesting to consider the issue if the Mannheim-Kazanas solution from [7] is the solution of CG equations of [9], and [10], that, as it was said before, is also conformal-invariant. Clarification of this issue is under consideration of our work here. We will show that the answer to this question is positive.

The specific feature of the result obtained lies in the fact that it broadens the possibilities for interpretation of the Mannheim-Kazanas solution. Actually, in case of CG the effective energy-impulse tensor is constructed on the basis of the Weyl vector, i.e. from the geometric characteristics of the space itself. Such energy-impulse tensor can be analyzed the same way as it is done in the GTR. The work provides the analysis that brings us to the conclusion about the abnormal character of thermodynamic properties of the produced geometrodynamic medium. One of the properties that confirm the abnormal character of the effective energy-impulse tensor is the negative character of the energy density at low enough values of the radial variable. In the end the results of the performed analysis are discussed.

## 2 MANNHEIM-KAZANAS SOLUTION

In the basis of quadratic conformal-invariant generalization of the general theory of relativity there is Weyl tensor $C_{\alpha\beta\mu\nu}$, that is defined through the Riemann tensor of curvature $R_{\alpha\beta\mu\nu}$, Ricci tensor $R_{\alpha\beta} = g^{\mu\nu}R_{\mu\alpha\nu\beta}$ and contracted Ricci tensor $R = g^{\mu\nu}R_{\mu\nu}$ in the following way:

$$C_{\alpha\beta\mu\nu} = R_{\alpha\beta\mu\nu} - \frac{1}{2}\left(g_{\alpha\mu}R_{\beta\nu} + g_{\beta\nu}R_{\alpha\mu} - g_{\alpha\nu}R_{\beta\mu} - g_{\beta\mu}R_{\alpha\nu}\right) + \frac{1}{6}R\left(g_{\alpha\mu}g_{\beta\nu} - g_{\alpha\nu}g_{\beta\mu}\right). \tag{2}$$

It is known from [7], [8] that if a scalar is used

$$L = C_{\alpha\beta\mu\nu}C^{\alpha\beta\mu\nu} \tag{3}$$

as a Lagrangian, a result of a standard variation procedure are Bach equations:

$$B_{\alpha\beta} = B^{(1)}_{\alpha\beta} + B^{(2)}_{\alpha\beta} = 0, \tag{4}$$

where

$$B^{(1)}_{\alpha\beta} = -R_{\alpha\beta}{}^{;\nu}{}_{;\nu} + R^{\nu}{}_{\alpha;\beta;\nu} + R^{\nu}{}_{\beta;\alpha;\nu} - \frac{2}{3}R_{;\alpha;\beta} + \frac{1}{6}g_{\alpha\beta}R^{;\nu}{}_{;\nu}, \qquad (5)$$

$$B^{(2)}_{\alpha\beta} = \frac{2}{3}RR_{\alpha\beta} - 2R_{\alpha\nu}R^{\nu}{}_{\beta} - \frac{1}{6}R^2 g_{\alpha\beta} + \frac{1}{2}g_{\alpha\beta}R_{\mu\nu}R^{\mu\nu}. \qquad (6)$$

Distinctive features of equations (4)-(6) are as follows:

(1) They are written only in terms of values $R_{\alpha\beta}$, $R$ and metric.

(2) The equations involve the forth order derivatives from the metric.

(3) The equations are invariant with regard to conformal transformations of the metric

$$g_{\alpha\beta} \to g^{*}{}_{\alpha\beta} = g_{\alpha\beta}e^{2\sigma\, x}. \qquad (7)$$

In Mannheim and Kazanas solution [7] the square of the interval is written in the form of

$$ds^2 = -B\, r\, dt^2 + \frac{dr^2}{B\, r} + r^2\left[d\theta^2 + \sin^2\theta d\varphi^2\right], \qquad (8)$$

and function $B\, r$ in the form of

$$B\, r\, =\, 1 - 3\beta\gamma\, -\frac{\beta\, 2-3\beta\gamma}{r} + \gamma r - \kappa r^2. \qquad (9)$$

Here, $\beta, \gamma, \kappa$ are some integration constants.

Solutions (8), (9) appeared to be fruitful from the point of view of explanation of flat galactic rotational curves and understanding of the dark matter nature. There are many publications on this issue, e.g., see [1]-[5].

## 3 EXACT SOLUTION FOR THE CG EQUATIONS

It is known that the static solution of GTR equations

$$R_{\alpha\beta} - \frac{1}{2}g_{\alpha\beta}R = \lambda g_{\alpha\beta} \qquad (10)$$

in case of center-symmetric static (CSS) problem is Schwarzschild solution in the space of de Sitter, that has the following form:

$$ds^2 = -e^{\gamma}dt^2 + e^{-\gamma}dr^2 + r^2\left[d\theta^2 + \sin^2\theta d\varphi^2\right], \qquad (11)$$

where

$$e^{\gamma} = 1 - \frac{r_0}{r} + \frac{\lambda_0}{3}r^2. \qquad (12)$$

Here, $r_0$ is gravitation radius, $\lambda_0 = \text{Const}$ - is lambda-term.

Equations (10) are a particular case of CG equations

$$R_{\alpha\beta} - \frac{1}{2}g_{\alpha\beta}R = -2A_{\alpha}A_{\beta} - g_{\alpha\beta}A^2 - 2g_{\alpha\beta}A^{\nu}{}_{;\nu} + A_{\alpha;\beta} + A_{\beta;\alpha} + \lambda g_{\alpha\beta} \qquad (13)$$

and are obtained from them if vector $A_{\alpha}$ is taken equal to zero. If we consider (11) as a solution of equation (13), then it should be written like this

$$ds^2 = -e^{\gamma} dt^2 + e^{-\gamma} dr^2 + r^2 \left[ d\theta^2 + \sin^2\theta d\varphi^2 \right],$$
$$e^{\gamma} = 1 - \frac{r_0}{r} + \frac{\lambda_0}{3} r^2, \quad A = 0, \quad \lambda = \lambda_0 = \text{Const.}$$
(14)

Let's presume that in case of CSS problem vector $A_\alpha$ has a radial component as the only component different from zero, which in (14) is denoted as $A$.

Equations (13) are invariant with regard to conformal conversions
$$g_{\alpha\beta} \to g^*_{\alpha\beta} = g_{\alpha\beta} e^{2\sigma},$$
$$A_\alpha \to A^*_\alpha = A_\alpha - \sigma_{,\alpha},$$
$$\lambda \to \lambda^* = \lambda e^{-2\sigma}.$$
(15)

In conformal conversions (15) not only metric in involved, but vector Weyl $A_\alpha$ and lambda-term $\lambda$ as well. So, if expressions (14) are transformed according to (15), then the transformed expressions will also be the solution of initial equation (13). Let's use this property and subject solution (14) to the conformal transformation with conformal factor $\exp\left[ 2\sigma(r) \right]$, where function $\sigma(r)$ depends only on the radial variable. The new solution will be again center-symmetric and static and have the following form:

$$ds^{*2} = -e^{\gamma(r)+2\sigma(r)} dt^2 + e^{-\gamma(r)+2\sigma(r)} dr^2 + e^{2\sigma(r)} r^2 \left[ d\theta^2 + \sin^2\theta d\varphi^2 \right],$$
$$A^* = -d\sigma(r)/dr, \quad \lambda^* = \lambda_0 e^{-2\sigma(r)}.$$
(16)

Then will go through coordinate transformation and replace radial coordinate $r$ with coordinate $R$ in such a way, that the new coordinate would be bright. That is let

$$R = r \cdot e^{\sigma(r)}.$$
(17)

The differentials of the old and new coordinates, as it comes from (17), are related through the ratio

$$dr = \frac{dR \cdot e^{-\sigma(r)}}{1 + r\sigma'(r)}.$$
(18)

After substitution (17), (18) in (16) we get

$$ds'^{*2} = -e^{2\sigma(r)} \left( 1 - \frac{r_0}{r} + \frac{\lambda_0}{3} r^2 \right) dt^2 + \frac{dR^2}{\left( 1 - \frac{r_0}{r} + \frac{\lambda_0}{3} r^2 \right) \left( 1 + r\sigma'(r) \right)^2} + R^2 \left[ d\theta^2 + \sin^2\theta d\varphi^2 \right],$$
$$A'^* = -d\sigma(r)/dr \frac{e^{-\sigma(r)}}{1 + r\sigma'(r)}, \quad \lambda'^* = \lambda_0 e^{-2\sigma(r)}.$$
(19)

In (19), term $r$ is to be considered as a function of a new radial variables $R$, i.e. as a function $r(R)$, determined from equations (18). On this stage function $\sigma(r)$ stays uncertain. Let's choose so that the following ratio would be true

$$g'^*_{00} g'^*_{11} = -1.$$
(20)

To meet ratio (20), as it comes from (19), the following ratio should be true

$$\frac{e^{2\sigma(r)}}{(1+r\sigma'(r))^2} = 1. \tag{21}$$

It is interesting to note that ratio (21) is obtained the same irrespective to what form the initial function $e^{\gamma(r)}$ was used. Then we take the square root from both parts (21) and get a differential equation to find function $\sigma(r)$ in the form

$$e^{-\sigma(r)}(1+r\sigma'(r)) = \eta, \tag{22}$$

where

$$\eta = \pm 1. \tag{23}$$

The solution of equations (22) is

$$e^{-\sigma(r)} = \left(\eta + \frac{r}{R_0}\right). \tag{24}$$

Here, $R_0$ is a constant with the dimension of the length. From the condition of the positive character of exponent $e^{\sigma(r)}$ we find that parameter $\eta$ should be equal to

$$\eta = +1, \tag{25}$$

and the value $R_0$ should be positive,

$$R_0 > 0. \tag{26}$$

As a result, we obtain

$$e^{\sigma(r)} = \frac{1}{\left(1+\dfrac{r}{R_0}\right)}. \tag{27}$$

With the account of (17) this results give the following relation between $r$ and $R$:

$$R = \frac{r}{\left(1+\dfrac{r}{R_0}\right)}, \quad r = \frac{R}{\left(1-\dfrac{R}{R_0}\right)}. \tag{28}$$

As a result, a conformal factor $e^{\sigma(r)}$ and expression for a new radial variable $R$ are found. Let's introduce a new symbol

$$F(R) = -g'^{*}_{00}. \tag{29}$$

Solution (19) will be written now like this

$$\left.\begin{array}{c} ds'^{*2} = -F(R)dt^2 + \dfrac{dR^2}{F(R)} + R^2\left[d\theta^2 + \sin^2\theta d\varphi^2\right], \\[6pt] A'^{*} = \dfrac{1}{R_0 - R}, \quad \lambda'^{*} = \dfrac{\lambda_0 R_0^2}{(R_0 - R)^2}. \end{array}\right\} \tag{30}$$

Here, $F(R)$ has the form

$$F(R) = \left(1-\frac{R}{R_0}\right)^2\left(1+\frac{r_0}{R_0} - \frac{r_0}{R}\right) + \frac{\lambda_0 R^2}{3}. \tag{31}$$

Let's consider particular cases.

<u>Schwarzschild solution</u>. We get it if in (30), (31) we have $\lambda_0 = 0$ and setting $R_0 \to \infty$.

<u>Kottler solution</u>. We get it if in (30), (31) setting $R_0 \to \infty$.

<u>Mannheim-Kazanas solution</u>. Squared interval in this solution has the form[1]

$$ds^2 = - e^{\gamma}{}_{MK} dt^2 + e^{-\gamma}{}_{MK} dR^2 + R^2 \left[ d\theta^2 + \sin^2\theta d\varphi^2 \right], \qquad (32)$$

where

$$e^{\gamma}{}_{MK} = 1 - 3\beta\gamma - \frac{\beta(2-3\beta\gamma)}{R} + \gamma R - \kappa R^2. \qquad (33)$$

If we take into account identity of function $e^{\gamma}{}_{MK}$ in (33) and function $F(R)$ in (31) and relate parameters $\beta, \gamma, \kappa$ with parameters $r_0, R_0$ in the following way:

$$\beta = \frac{r_0}{R_0 \left(2 + 3\frac{r_0}{R_0}\right)}, \quad \gamma = -\frac{\left(2 + 3\frac{r_0}{R_0}\right)}{R_0}, \quad \kappa = -\frac{1}{R_0^2}\left(1 + \frac{r_0}{R_0}\right) + \frac{\lambda_0}{3}, \right\} \qquad (34)$$

then it will become evident that (30), (31) is the solution from the class of Mannheim-Kazanas solutions. Actually:

* Both solutions relate to the class of center-symmetrical static solutions, where bright radial variable is implemented.
* 00-component of metric in both solutions comprise the terms with degrees $R^{-1}, R^0, R^1, R^2$.
* Equalization of the coefficients at the same degrees of radial variable result into four ratios. We get the relation (34) between the parameters from three ratios. The forth ratio is met automatically.

## 4 PROPERTIES OF GEOMETRODYNAMICS MEDIUM

From the point of view of GTR equations solution (30), (31) is the solution with a non-zero energy-impulse tensor. Components of tensor $T_{\alpha\beta}$ different from zero can be found from GTR equations

$$R_{\alpha}^{\beta} - \frac{1}{2}\delta_{\alpha}^{\beta} R = T_{\alpha}^{\beta}, \qquad (35)$$

written for static center-symmetric case (see [13], [14], for example). These equations result into the following formulas:

$$T_0^0 = e^{\gamma}\left[\frac{\gamma'}{R} + \frac{1}{R^2}\right] - \frac{1}{R^2}, \qquad (36)$$

$$T_1^1 = e^{\gamma}\left[\frac{\gamma'}{R} + \frac{1}{R^2}\right] - \frac{1}{R^2}, \qquad (37)$$

---

[1] Radial variable, which is marked as $r$ in Mannheim-Kazanas solution (8), (9), here is marked as $R$.

$$T_2^2 = e^\gamma \left[ \frac{\gamma''}{2} + \frac{\gamma'^2}{2} + \frac{\gamma'}{R} \right]. \tag{38}$$

Energy-impulse tensor components are determined only through function $e^\gamma$, as it comes from (36)-(38). Using formulas (30), (31) we get:

$$\left. \begin{array}{l} -U = T_0^0 = \dfrac{3}{R_0^2}\left(1 + \dfrac{r_0}{R_0}\right) - \dfrac{2}{R_0}\left(2 + 3\dfrac{r_0}{R_0}\right)\dfrac{1}{R} + 3\dfrac{r_0}{R_0}\dfrac{1}{R^2} + \lambda_0 \\[6pt] P_R = T_1^1 = -U, \\[6pt] P_\theta = T_2^2 = \dfrac{3}{R_0^2}\left(1 + \dfrac{r_0}{R_0}\right) - \dfrac{1}{R_0}\left(2 + 3\dfrac{r_0}{R_0}\right)\dfrac{1}{R} + \lambda_0. \end{array} \right\} \tag{39}$$

At low values of radial variable geometrodynamic medium described with tensor $T_{\alpha\beta}$ (39) has the following abnormal features in equations of state:

- negative density of energy $U < 0$,
- not equal between themselves stress tensor components $P_R \neq P_\theta$,
- radial pressure is equal to the energy density with the opposite sign, $P_R = -U$.

## 5 DISCUSSION

The basic results of the work can be summarized as follows.

∗ It is shown that the center-symmetric static Mannheim-Kazanas solution is the solution of not only Bach equations (4)-(6), but the solution of equations of conformal geometrodynamics (13).

∗ It comes from CG equations that the scalar field contained in the conformal factor brings a non-zero energy-impulse tensor with components (36)-(38).

∗ Analysis of tensor (36)-(38) shows that at small values of radial variable geometrodynamic medium for this CG solution has abnormal equations of state: negative energy density $U < 0$, non-equal components of stress tensor $P_r \neq P_\theta$, but the radial pressure is positive and modulo is equal to the energy density, $P_r = -U$.

In references [1]-[5], [15] they note that within standard GTR it is impossible to explain numerous observation data that refer to the flat galactic rotational curves without attraction of a dark matter hypothesis. Dark matter carriers are not found yet, and this makes it important to look for an alternative explanation of the observed data. Natural explanation for these data is obtained within Mannheim-Kazanas solution (8), (9). But this solution is not the solution of GTR equations in the standard form. At the same time it is both the forth order solution of Bach equations and the second order solution of CG equations (that has conformal invariance and is related to the Weyl space). Bach equations [16] obtained from quadratic action by Weyl tensor are the most known variant of conformal invariant generalization of the GTR equations. CG equations on the basis of Weyl geometry [9], [10] are less used as an alternative for GTR (however note [17], [18]), but they are the simplest variant of realization of conformal invariance, as there generalization is done by introduction of a so called gauge vector field $A_\alpha$ into GTR equations when preserving the order of the obtained differential equation. This procedure is analogous to the one that is used in physics when gauge fields are introduced. CG equations produced in this case allow the analysis of the properties of the geometrodynamic medium that depend on the compensating vector field, in the terms of some effective energy-impulse tensor.

Substitutions of GTR equations with CG equations [10] results to the transition from Riemann space to space of Weyl, and this requires reconsideration of the conceptual properties of space. The space stays 4D with signature $-+++$. But in space of Weyl it is necessary to register not only the reference frame, but also the scale of length (time). This procedure is analogous to the calibration of the vector-potential in Maxwell electrodynamics. The selection of a single in space gauge corresponds to the reversal into zero for vector $A_\alpha$. Mannheim-Kazanas solution takes here the form of Schwarzschild solution. Another choice of selected gauge length is possible, when the square of the interval has general expression (14). This choice unambiguously leads to Mannheim-Kazanas solution (8),(9). Dependency of the interpretation of the solution of CG equations on the choice of calibration of vector $A_\alpha$ in space of Weyl has a general character. It is possible to presume that we can neglect this dependency in the range of lengths from the ones of characteristic elementary particles to the sizes of star systems and use a constant gauge, i.e. assume $A_\alpha \approx 0$. But starting with galactic distances the situation changes, which is indicated by flat galactic rotational curves. We may make a conclusion that if not involving the dark matter concept the reality in galactic scales is better described not with GTR equations in their standard form, but with the gravitation equations in conformal invariant form.